\documentclass[hyper,11pt,letterpaper]{JHEP3}
\usepackage[dvips]{epsfig}
\usepackage{amsmath,amssymb,epsf,amsfonts}
\usepackage{graphicx}
\usepackage{dcolumn}
\usepackage{bm}

\def\<{\langle}

\def\>{\rangle}
\def\({\left (}
\def\){\right )}
\def\[{\left[}
\def\]{\right]}

\def\beq{\begin{equation}}
\def\eeq{\end{equation}}

\newcommand{\bea}{\begin{eqnarray}}
\newcommand{\eea}{\end{eqnarray}}

\title{A Holographic Quantum Critical Point at Finite Magnetic Field
and Finite Density}

\author{ Kristan Jensen, Andreas Karch,
and Ethan G. Thompson
\vspace{0.4cm} \\
\small  Department of Physics, University of Washington,
Seattle, WA 98915-1560, USA\\
\small  ~\,E-mail: \tt{kristanj,akarch,egthomps@u.washington.edu}
}

\maketitle

\abstract{
We analyze the phase diagram of ${\cal N}=4$ supersymmetric Yang-Mills theory
with fundamental matter in the presence of a background magnetic
field and nonzero baryon number. We identify an isolated quantum critical point separating
two differently ordered finite density phases.  The ingredients that give rise to this transition are generic in a holographic setup, leading us to conjecture that such critical points should be rather common.  In this case, the quantum phase transition 
is second order with mean-field exponents.  We characterize the neighborhood of the critical point at small temperatures and identify some signatures of a new phase dominated by the critical point.  We also identify the line of transitions between the finite density and zero density phases.  The line is completely determined by the mass of the lightest charged quasiparticle at zero density.  Finally, we measure the magnetic susceptibility and find hints of fermion condensation at large magnetic field.
}
\preprint{INT-PUB-10-010}
\begin{document}

\section{Introduction}

Quantum phase transitions are continuous phase transitions at zero temperature \cite{sachdevbook}.  The transitions occur at quantum critical points, which are reached when a set of control parameters are properly tuned.  These quantum critical points are an active area of research in the condensed matter community.  For example, they play an important role in a candidate theory of high-$T_c$ superconductors \cite{sachdev-2009}. Often, one is interested not just in the effective theory at the critical point, but also in a neighborhood of the critical point.  In this paper, we study an example of a quantum critical point in the context of holography.

Holography establishes a large class of strongly-coupled, scale-invariant theories which can be solved using a dual gravitational description.  This duality has recently found many applications to condensed matter physics.  Reviews of this work can be found in \cite{Hartnoll:2009sz,Herzog:2009xv,McGreevy:2009xe}.  In principle, any such scale-invariant theory is potentially the low-energy description of a theory at a quantum critical point. However, to our knowledge, no scale-invariant theory with a gravitational dual has been shown to be a quantum critical point between two differently-ordered, finite-density phases arising from tuning the control parameters of a microscopic model. In this work, we give an explicit realization of such a system.  We construct and analyze a holographic setup which exhibits an isolated quantum critical point in its phase diagram as a function of magnetic field and charge-carrier density.

Two complementary approaches to realize holographic finite density systems have been developed. ``Bottom-up" models study an effective gravitational theory in the background of a charged black hole in an attempt to simulate a strongly-coupled theory at finite density.  By picking and choosing the ingredients and interactions of the effective theory, holographic examples of diverse condensed matter phenomena have been constructed, including superconductivity and -fluidity \cite{Gubser:2008px,Hartnoll:2008vx} as well as Fermi- \cite{Cubrovic:2009ye} and non Fermi-liquidity \cite{lee-2009-79,Liu:2009dm,Faulkner:2009wj}. The only drawback of this approach is that typically one does not have a clear understanding of what the dual field theory is. Nonetheless, these models see an emergent ``quantum criticality'': the bottom of the geometry has an AdS$_2$ factor (or, after backreaction, a Lifshitz-like space \cite{Hartnoll:2009ns}), and so the dual theory appears to have an emergent $0+1$ dimensional description in the IR. Even in cases where a scalar condensate is formed one typically sees a new AdS geometry emerge in the infrared \cite{Gubser:2009cg}. However, these systems are already tuned \emph{to} criticality from the start.  It is then unclear not only what dual theory one is working with in these setups, but also what parameters are being adjusted to reach criticality.

An alternative ``top-down" approach is to start with a known gauge/gravity pair with at least a global $U(1)$ symmetry and to turn on a chemical potential for that $U(1)$. Knowledge of the explicit field theory Lagrangian, in particular the knowledge gained from any weak-coupling limit it has in addition to the strong-coupling limit in which the gravitational description is valid, is an important guide to understanding the phenomena one sees. The main disadvantage of this approach is that, while one is studying a known and consistent field theory, one has very little control over what dynamical behavior dominates the low energy physics of the system. In particular, isolated quantum critical points at finite density have been difficult to realize in this framework.

An important tool to build ``top-down" models is the use of probe flavor branes \cite{Karch:2000gx,Karch:2002sh}. On the field theory side one adds a finite number $N_f$ of fundamental representation charge carriers (the ``electrons") to a known gauge theory with a large number of colors $N_c$ and a gravitational dual. This introduces a new $U(1)$ global symmetry, ``baryon"  or ``electron" number, under which only the new fields are charged. For $N_f/N_c \ll 1$, the dynamics of the new charge carriers do not backreact on the underlying field theory (the ``phonon bath"); transport phenomena associated with electron number are dominated by electrons strongly interacting with the underlying phonon bath. The addition of the extra charge carriers is implemented in the gravity dual by including $N_f$ flavor D-branes
 which do not backreact on the dual geometry in the probe limit $N_f/N_c \ll 1$. In addition, the flavor D-branes support a $U(1)$ gauge field on their wordvolume whose dynamics is universally governed by the Dirac-Born-Infeld (DBI) action. The transport properties associated with electron number on the field theory side can be found by studying the classical dynamics of this DBI gauge field.

\begin{figure}
{

\begin{center}
\includegraphics[scale=.73]{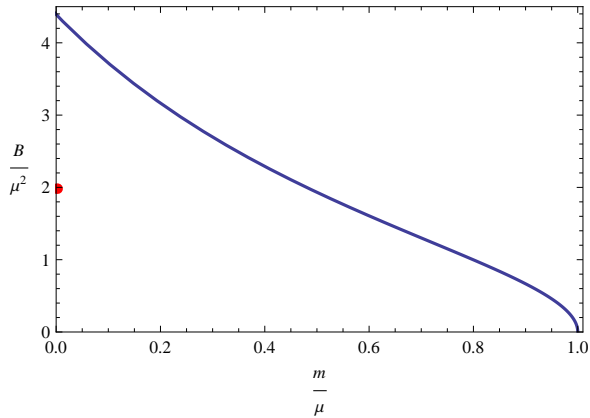}
\vspace*{-10pt}
\caption
  {\label{d3d7phases}The phase diagram for the D3/D7 system in the $m/\mu$ - $B/\mu^2$ plane at zero temperature. The red dot indicates a critical point at which we go from the high-density ``boson-dominated" phase and no condensate to
  the low-density, presumably ``fermion-dominated" phase. The blue line indicates a phase transition to a zero density phase with properties identical to the vacuum at nonzero magnetic field, located in the upper right hand side of the diagram. While we can not rule out the possibility that this transition is first order away from the $B=0$ axis, our numerics suggest that it is a second or higher order transition throughout.
}
\vspace*{-10pt}
\end{center}
}
\end{figure}

The study of finite density in flavor brane systems, and in particular the simplest D3/D7 system of \cite{Karch:2002sh}, was initiated in  \cite{Kobayashi:2006sb}\footnote{See \cite{Apreda:2005yz,Kim:2006gp,Horigome:2006xu,Parnachev:2006ev} for earlier work on finite density flavor brane systems.}. Qualitatively, one finds that the finite density attracts the flavor branes to the horizon of the bulk spacetime. While at zero density both horizon-crossing probe brane embeddings as
well as embeddings that always stay above the horizon can be found \cite{Babington:2003vm}, all consistent finite density brane embeddings cross the horizon. The field theory density is realized by a bulk electric field, sourced by charge located behind the horizon.

In this work we generalize these studies by exposing the finite density of matter to an external magnetic field. One important result we are able to establish is the existence of an isolated critical point in the phase diagram of the D3/D7 system at zero temperature as a function of magnetic field, density and a ``coupling constant", the bare mass of the charge carriers in the underlying Lagrangian. The phase diagram of this D3/D7 system, which constitutes the main result of our paper, is displayed in Fig. \ref{d3d7phases}.   While first order phase transitions in probe brane systems are generic \cite{Mateos:2006nu}, one needs to work a little harder to find continuous transitions. Continuous transitions with calculable critical exponents can be found when studying a theory with probe branes at finite volume \cite{Karch:2009ph}. One example of an analytically-tractable second-order transition involving finite density is the D3/D7 system at zero magnetic field, as exhibited in \cite{Karch:2007br}. The lightest charged quasiparticles of that theory generally have some finite mass; a chemical potential larger than that mass is required to populate a nonzero charge density. At the critical chemical potential one therefore expects a phase transition from a trivial zero-density phase (equivalent to the vacuum) to a finite-density phase. In nature, these phase transitions are often first order, but in the D3/D7 system the transition is second order \cite{Karch:2007br}. In fact, we find an analogue of this phase transition for all values of the magnetic field as indicated by the blue line in Fig. \ref{d3d7phases}. Our numerics indicate that the transition is at least second order everywhere along the line\footnote{This has to be contrasted with the transition in the $(\mu, T)$-plane, where the second order transition of \cite{Karch:2007br} at zero temperature turns into a third order and eventually a first order transition \cite{Faulkner:2008hm}.}. The critical chemical potential varies as a function of the magnetic field. We find it to be the mass of the lightest charged quasiparticle (as a function of the magnetic field). Although this is a continuous phase transition at zero temperature, it is still qualitatively different from the typical quantum critical points of interest in the condensed matter community. Instead of describing a critical point that results from a competition between two different ordered phases in a finite density of material, the second order transition of \cite{Karch:2007br} simply describes a transition from the trivial vacuum to a phase where there is actually something. It is a property only of the grand canonical ensemble and is obviously absent in the canonical description.
To our knowledge, the quantum critical point we find is the first example of a continuous transition between two finite density phases with different order in a theory with a gravitational dual \footnote{While we were finishing this work, the authors of \cite{Evans:2010iy} released a paper that studies the phase diagram of the D3/D7 system with a magnetic field, density, and temperature.  Their zero temperature results agree with ours.}.

The phase transition we describe arises from a competition between finite density and magnetic fields when the ``electrons'' have zero bare mass. In the gravitational system, nonzero density pulls the brane to the horizon, whereas a magnetic field repels it. From the field theory point of view, it has been well known that a magnetic field at zero density triggers chiral symmetry breaking \cite{Filev:2007gb}, presumably due to the interactions of fermion spin with the magnetic field. This chiral condensate also serves as the order parameter that detects our transition. In contrast, at finite density and zero magnetic field one finds the boson-dominated chirally symmetric phase described in \cite{Karch:2007br}. As we will argue in our concluding section, we would like to interpret our phase transition as a competition between bosonic and fermionic condensates.

One important point to note is that while our numeric calculations are particular to the D3/D7 system, the ingredients are very general. For a generic probe system with an effective AdS wordvolume a very similar phase transition should arise. Most of our analysis relies on studying the DBI action in AdS$_5$ and does not make any reference to the internal geometry. The mass of the slipping mode (that is, the dimension of the fermionic bilinear condensate) is the only geometric input into the DBI action that influences the location of the critical point. This large class of new quantum critical points should be a very useful starting point for future investigations.  The analogous phase diagram for the D3/D5 system of \cite{Karch:2000gx} was recently studied in \cite{Wapler:2009rf}. However, the analogue of the quantum critical point we exhibit was missed in this work since only the chirally symmetric embedding was considered for the massless ``electron'' case.  We have identified a quantum critical point in the D3/D5 system as well.  This critical point exhibits some novel properties and will be described in a forthcoming paper in conjunction with Dam Son \cite{D3D5}.

The outline of the paper is as follows. In the following section we will discuss the theory with massless charge carriers, exhibiting the quantum critical point that arises from the competition of magnetic field and finite density.
In Sec. 3, we then map out the full phase diagram at zero temperature as a function of charge carrier mass and elucidate the second order phase transition between the zero density and finite density phases. In our concluding section, Sec. 4, we characterize an anomalous phase of matter and present a model that explains the phenomena we see in terms of a competition between bosons and fermions. We also make a conjecture regarding the generality of similar critical points.  We present the details of our numerical methods in the appendix.

\section{A quantum phase transition for massless flavor}
\subsection{The setup and action}\label{setup}

In this work, we study $N_f$ flavors of $\mathcal{N} = 2$ supersymmetric matter coupled to an $SU(N_c)$ $\mathcal{N} = 4$ super Yang-Mills (SYM) gauge theory in the strong 't Hooft coupling $\lambda$, large $N_c$, and quenched $N_f\ll N_c$ limits.  The dual gravitational description of this system is that of $N_f$ D7 branes probing the near-horizon $AdS_5 \times S^5$ geometry of $N_c$ D3 branes \cite{Karch:2002sh}.  Without turning on non-abelian field strengths, the DBI action of the probe D7 branes is
\begin{equation}
S_{DBI} = - N_f T_{D7} \int d^8 \xi \sqrt{ - \textrm{P}[g] + F}.
\end{equation}
Here, $T_{D7}$ is the tension of the D7 branes, $\text{P}[g]$ is the induced metric on the branes, $F$ is the field strength of the diagonal $U(1)$ worldvolume gauge field living on them (we have absorbed a factor of $2 \pi \alpha'$ into our definition of the field strength, leaving it dimensionless), and $\xi^A$ indexes worldvolume coordinates.  For the embeddings we consider, the Chern-Simons terms in the D7 action vanish and the brane action \emph{is} the DBI part.

We find it convenient to write the metric of the background geometry as \cite{Kruczenski:2003be}
\begin{equation}
\label{backgroundmetric}
g = \frac{r^2 + y^2}{R^2} \left( -(dx^0)^2 + d\vec{x}^2 \right) + \frac{R^2}{r^2 + y^2} \left( dr^2 + r^2 d\Omega_3^2 + dy^2 + y^2 d\phi^2 \right),
\end{equation}
where $R$ is the radius of both the $AdS$ space and the five-sphere and we have broken up the six dimensions transverse to the D3 branes into a four-dimensional and a two-dimensional subspace, writing both spaces in polar coordinates. In these coordinates, the boundary of AdS occurs when either $r\rightarrow\infty$ or $y\rightarrow \infty$ and the Poinc\'are horizon is located at $y=r=0$.

We take an ansatz that the D7 branes wrap the $AdS$ space as well as the three-sphere in the metric above.  Additionally, the position of the branes in the $y$ direction is allowed to depend on the ``radial coordinate'' $r$ and we consider solutions with constant $\phi$.  We use the $U(1)$ symmetry that rotates $\phi$ to set it to zero.
The embedding is then described by one function $y(r)$.
With this ansatz the induced metric
\begin{equation}
\text{P}[g]=G_{AB} d\xi^A d\xi^B, \,\,\,\,\,\,\,\,\, G_{AB} = \frac{ \partial X^a}{\partial \xi^A} \frac{\partial X^b}{\partial \xi^B} g_{ab},
\end{equation}
where $a, b$ run over all ten dimensions, $g_{ab}$ is the background metric (\ref{backgroundmetric}), and the $X^a$ are the embedding functions, is given by
\begin{equation}
\text{P}[g] = \frac{r^2 + y^2}{R^2} \left( -(dx^0)^2 + d\vec{x}^2 \right) + \frac{R^2}{r^2 + y^2} \left( \left (  1+ (y')^2 \right ) dr^2 + r^2 d\Omega_3^2  \right).
\end{equation}

We pause to note that these embeddings preserve the $SO(4)$ isometry of the three-sphere and, when the embedding is simply $y=0$, the $U(1)$ isometry that rotates $\phi$.  These isometries are mapped onto R-symmetries of the dual field theory. In particular, the $U(1)_{\rm R}$ symmetry is a chiral symmetry that is explicitly broken by mass terms for the flavor multiplet and spontaneously broken when the operator dual to the field $y$ attains a vev.  We will come back to this later.

We now turn on a finite density of matter and a magnetic field in the dual field theory, using the diagonal $U(1)$ flavor symmetry representing electron number.  On the gravity side, the conserved current for this $U(1)$ is dual to the diagonal $U(1)$ subgroup of the worldvolume gauge theory on the branes.  The field theory density is dual to a radial electric field on the gravity side, $F=A'_0(r)dx^0\wedge dr$, sourced by bulk charge behind the AdS horizon.  Our branes therefore cross the AdS horizon.  We also turn on a background magnetic field by introducing a constant field strength $F=B \,dx^1\wedge dx^2$ in both the field theory and the bulk.

Plugging in these ans\"atze for the induced metric and gauge field into the DBI action, integrating over both the internal three-sphere and the non-compact $\mathbb{R}^4$, and representing the brane tension $T_{D7}$ and $\alpha'$ in field theory quantities,
\begin{equation}
T_{D7} = \frac{1}{g_s (2\pi)^3 (2\pi\alpha')^4},~~~ g_s = \frac{ \lambda}{4 \pi N_c},~~~ (\alpha')^4 = \frac{R^8}{\lambda^2},
\end{equation}
we arrive at the following action governing the dynamics of the embedding
scalar $y(r)$ and the worldvolume gauge fields,
\begin{equation}
S_{D7} = - \frac{ N_c N_f \lambda}{ (2 \pi )^4 } \frac{ V_4}{R^4} \int \frac{dr}{R} \frac{ r^3}{ R^3 }  \sqrt{ \left( 1 + y'(r)^2 -  A'_0(r)^2 \right) \left(1 + \frac{B^2R^4}{(r^2 + y(r)^2)^2} \right)},
\end{equation}
where $V_4$ is the volume of the $\mathbb{R}^4$.

Finally, we rescale all of our coordinates by a factor of $R$, divide the action by $V_4$ to give the action density (abusing notation by not changing symbols), and define the overall normalization to be $\mathcal{N} = \frac{N_f N_c \lambda}{(2 \pi)^4 }$, leaving us with the action density
\begin{equation}
\label{originalAction}
S_{D7} = - \mathcal{N} \int dr\, r^3 \sqrt{ \left( 1 + y'(r)^2 -  A'_0(r)^2 \right) \left(  1 + \frac{B^2}{(r^2 + y(r)^2)^2} \right)}.
\end{equation}

From this action, we could derive the coupled equations of motion for the fields $A_0(r)$ and $y(r)$.  However, because the action only depends on the derivative of $A_0$, we replace $A_0$ in favor of the conserved quantity $d\equiv \frac{\delta\mathcal{L}_{D7}}{\delta A'_0(r,x)}$. This \emph{is} the electron density of the dual theory \cite{Kobayashi:2006sb},
\begin{equation}
d = \lim_{\Lambda\rightarrow\infty}\frac{ \delta S_{D7} }{ \delta A_0(r=\Lambda,x) } = \mathcal{N} r^3 A'_0(r) \sqrt{  \frac{   1 + \frac{B^2}{(r^2 + y(r)^2)^2} }{ 1 + y'(r)^2 -  A'_0(r)^2 } }.
\end{equation}
Next, we solve for $A'_0$ in terms of $y$ and $d$, giving
\begin{equation}
A'_0(r)^2 = \frac{ d^2 (1 + y'(r)^2) }{ (d^2 + \mathcal{N}^2 r^6)  +\frac{\mathcal{N}^2 r^6 B^2}{(r^2 + y(r)^2 )^2 } }.
\end{equation}
We now Legendre transform to eliminate $A_0$ and find the action at fixed rescaled density $d=\mathcal{N}\rho$,
\begin{eqnarray}
\label{canonS}
\hat{S}_{D7}[d;y,y']  & = & S_{D7}[ A'_0(d,y,y'); y, y'] - \int dr\, \frac{\delta \mathcal{L}_{D7}}{\delta A'_0}A'_0(d; y, y') \nonumber \\
& = & - \mathcal{N}\int dr  \sqrt{ 1 + y'(r)^2 } \sqrt{ \rho^2 + r^6 + \frac{r^6 B^2}{(r^2 + y(r)^2 )^2} } .
\end{eqnarray}
From $\hat{S}_{D7}$ we can derive a single nonlinear equation of motion for $y(r)$ in terms of the density $\rho$ and the magnetic field $B$.
The resulting second-order differential equation for $y(r)$ has
two singular points, one at the boundary $r \rightarrow \infty$ and one at $r = 0$.  We can perform a Frobenius expansion at either singular point and from there determine suitable boundary conditions on the field $y$.

Our finite density embeddings cross the AdS horizon as described in \cite{Kobayashi:2006sb}.  These embeddings therefore obey the boundary condition $y(0)=0$.  The Frobenius expansion near $r = 0$ then takes the form
\begin{equation}
\label{horFrob}
y(r) = \gamma_1 r + \sum_{i = 2}^\infty \gamma_i r^{i},
\end{equation}
where $\gamma_1$ is a free parameter, and all the other $\gamma_i$ are determined by $\gamma_0$.

The Frobenius expansion near the boundary takes the form
\begin{equation}
\label{bdyFrob}
y(r) = c_0 + \frac{c_2}{r^2} + \sum_{i = 2}^\infty \frac{c_{2i}}{r^{2i} },
\end{equation}
where both $c_0$ and $c_2$ are free parameters and the higher $c_i$ are determined by them.  After performing holographic renormalization \cite{deHaro:2000xn}, which we review in Sec. \ref{holorg}, we learn that $c_0$ is essentially the bare mass, $m$, for the $\mathcal{N} = 2$ matter.  More precisely, it is the source for an operator that includes
the standard chiral condensate bilinear in the matter fermions and
whose specific form is given in \cite{Kobayashi:2006sb}.  The bulk field $y$ is dual to this operator and its expectation value (divided by $\mathcal{N}$) is given by $c =-2c_2$, found by
\begin{equation}
\left\langle \mathcal{O}_y\right\rangle=\mathcal{N}c=\lim_{\Lambda\rightarrow\infty}\frac{1}{\Lambda^3}\frac{\delta \hat{S}_{D7}}{\delta y(r=\Lambda,x)}.
\end{equation}
This condensate will serve as an order parameter for chiral symmetry breaking.

Except for the trivial solution $y=0$, the equation of motion for $y$ can only be solved numerically.  We therefore elect to shoot; we can shoot from the bottom of the brane at $r=0$ by dialing $\gamma_1$ in the near-horizon solution Eq. (\ref{horFrob}).  Each solution will correspond to an extremum of the dual theory at the mass and condensate represented in the large $r$ behavior of that solution.  Alternatively, we could impose a boundary condition at large $r$ -- say $m=0$ -- and dial the condensate until the embedding crosses the horizon, corresponding to an extremum of the massless theory.

\subsection{Holographic Renormalization}
\label{holorg}
The exponentiated (negative) on-shell action of the bulk theory is the generating functional of the dual field theory \cite{Witten:1998qj, Gubser:1998bc}.  The on-shell action is then (minus) the free energy of the dual state.  In our case, the bulk theory is the dimensional reduction of type IIB supergravity on AdS$_5\times S^5$ onto AdS$_5$ plus $N_f$ D7 branes.  The action of the supergravity fields is much larger than the DBI action of the branes.  The former corresponds to an order $N_c^2$ contribution to the free energy of the field theory, while the DBI action of the probe branes gives an order $N_c$ contribution.  It is this order $N_c$ term (the leading flavor contribution) that we consider hereafter.  Moreover, since the on-shell action is extremized on solutions $y(r)$, a given solution corresponds to an extremum of the dual field theory.

The DBI action at fixed density naturally corresponds to the free energy at fixed density, i.e. in the canonical ensemble,
\begin{equation}
F(B, d, m) = -\hat{S}_{D7} |_{on-shell}.
\end{equation}
However, simply plugging bulk solutions into the DBI action and attempting to integrate leads to divergences.  The same problem usually emerges when differentiating the action to obtain expectation values in the dual field theory.  These divergences are well understood and can be removed by holographically renormalizing the bulk theory \cite{deHaro:2000xn,Karch:2005ms}.  On the bulk side, the gravitational theory naively diverges because it lives in an infinite volume.  These bulk divergences occur near the AdS boundary and correspond precisely to the UV divergences of the dual theory.  Holographic renormalization is the process of regulating a bulk theory in a diffeomorphism invariant fashion, leading to both a well-defined bulk theory and its dual.

The easiest way to perform holographic renormalization is to examine the near-boundary behaviour of the action density for a general solution.  The action will have a number of divergences, each of which will be removed with an appropriate counterterm.  Using the near-boundary solution, Eq. (\ref{bdyFrob}), for $y(r)$ we find the near-boundary expansion of the action,
\begin{equation}
\hat{S}_{D7} = -\mathcal{N}\int drd^4x \left( r^3 + \frac{B^2}{2r} + \mathcal{O}(r^{-2}) \right).
\end{equation}
The action has two divergences.  The first term corresponds to the infinite volume of an asymptotically AdS$_5$ space, while the second term leads to a logarithmic divergence and corresponds to a conformal anomaly of the dual theory.  We proceed to regulate the action by cutting off the integral at some large value of $r$, which we call $\Lambda$.  The divergent parts of the action are
\begin{equation}
\hat{S}_{D7} \supset - \mathcal{N}\int d^4x\left(\frac{\Lambda^4}{4} + \frac{B^2}{2} \log \Lambda\right).
\end{equation}
We regulate these divergences by adding counterterms that live purely on this fixed-$r$ slice and are invariant under diffeomorphisms on the slice\footnote{This is a radial slice of the asymptotic AdS$_5$ factor in the branes, rather than in the full eight-dimensional worldvolume.}.
In our case, we need to add the counterterms
\begin{equation}
S_{\rm counter}=\mathcal{N} \int_{r = \Lambda} d^4x \frac{ \sqrt{ -\gamma}}{4} ( 1 + F_{\mu\nu} F^{\mu\nu} \log \Lambda ),
\end{equation}
where $\gamma$ is the induced metric on the $r = \Lambda$ slice and the indices $\mu,\nu$ on the field strength run only over the slice coordinates $\mu=0-3$.  These terms exactly cancel the near-boundary divergences of the bulk theory.  Adding the counterterms and taking the $\Lambda\rightarrow\infty$ limit, we obtain the regulated, diffeomorphism-invariant action
\begin{equation}
\label{renS}
\hat{S}_{D7,ren}[d;y,y']\equiv \lim_{\Lambda\rightarrow\infty}\left[ \int^{\Lambda}dr \, \hat{\mathcal{L}}_{D7}[d;y,y']+ S_{\rm counter}\right].
\end{equation}
It is this renormalized action that corresponds to the renormalized generating functional of the dual theory.

As alluded to above, the logarithmic counterterm indicates a conformal anomaly of the dual theory \cite{Henningson:1998gx,Graham:1999pm}.  To see this, recall that dilatations of the field theory correspond to bulk scale transformations
\begin{equation}
r' = \lambda r,~~~ (x^{\mu})' = \lambda^{-1} x^\mu.
\end{equation}
A bulk field dual to an operator of dimension $\Delta$ transforms with a power of $\lambda^\Delta$, and so the magnetic field transforms as $B'=\lambda^2 B$. The original DBI action is invariant, but the logarithmic counterterm is not.  Writing it in terms of the new primed coordinates we find
\begin{equation}
S_{\rm counter}' = \mathcal{N}\int d^4x' \frac{ (\Lambda')^4}{4} \left( 1 + \frac{ 2 (B')^2}{(\Lambda')^4} \log \,\Lambda' \right) = S_{\rm counter} + \mathcal{N}\int_{r=\Lambda} d^4 x \sqrt{-\gamma}\frac{ F_{\mu\nu}F^{\mu\nu} \log \lambda}{4},
\end{equation}
where $\Lambda'=\lambda \Lambda$ is the new cutoff.  With this result, the renormalized free energy then transforms as
\begin{equation}
\label{confAnom}
F_{ren}'= F_{ren}-\mathcal{N}\int d^4x \frac{F_{\mu\nu}F^{\mu\nu}}{4}\log\, \lambda,
\end{equation}
written in terms of the field strength in the field theory and the Minkowski metric.  The dilatation invariance of the free energy of the field theory is violated, corresponding to a conformal anomaly proportional to $F^2$.

Finally, we need to say a few words about our renormalization scheme.  By dimensional analysis, we could also include a counterterm of the form
\begin{equation}
S_{\rm finite}=\alpha\int_{r=\Lambda} d^4x \sqrt{-\gamma}F_{\mu\nu}F^{\mu\nu},
\end{equation}
in our renormalized action Eq. (\ref{renS}).  This term introduces no divergences for any value of $\alpha$, but rather shifts the on-shell action by the amount $\alpha B^2$.  Different values of $\alpha$ correspond to different renormalization schemes for the dual theory, where some correlation functions will contain scheme-dependent terms.  Since we must make a choice of scheme, we have decided to choose the simplest scheme $\alpha=0$ in this work.

\subsection{Numerical Analysis of Phase Structure at $T = 0$, $m = 0$}
\label{critPoint}
We now investigate the phase diagram of the D3/D7 system at zero bare mass and zero temperature.  Fixing the density $\rho$, we use the shooting techniques discussed in Sec. \ref{setup} to find the brane embeddings that correspond to the theory at zero mass as we dial the magnetic field.   We find two distinct regimes, separated by a critical magnetic field, which we found to be $B_c/\rho^{2/3} = 2.1387341...$  For completeness, we write this critical magnetic field in terms of the physical charge density and field theory quantities as
\begin{equation}
B_c=\beta \frac{ (2 \pi)^{ 8/3} d^{2/3}}{ (N_c N_f)^{2/3} \lambda^{2/3} },
\end{equation}
where $\beta$ is the numerical factor above.

For small $B < B_c$ (the ``high density'' phase), there is only one solution that corresponds to zero mass, the trivial embedding $y=0$.  The condensate vanishes and this phase is chirally symmetric.  However, for $B > B_c$ (the ``large magnetic field'' phase), we find at least two solutions corresponding to zero mass.  There is now a second, non-trivial solution that corresponds to a nonzero condensate.  We calculated the renormalized free energies for both solutions and found that the non-trivial embedding always has a lower free energy and thus is the thermodynamically preferred equilibrium state of the theory.

\begin{figure}
\begin{center}
\includegraphics[scale=0.7]{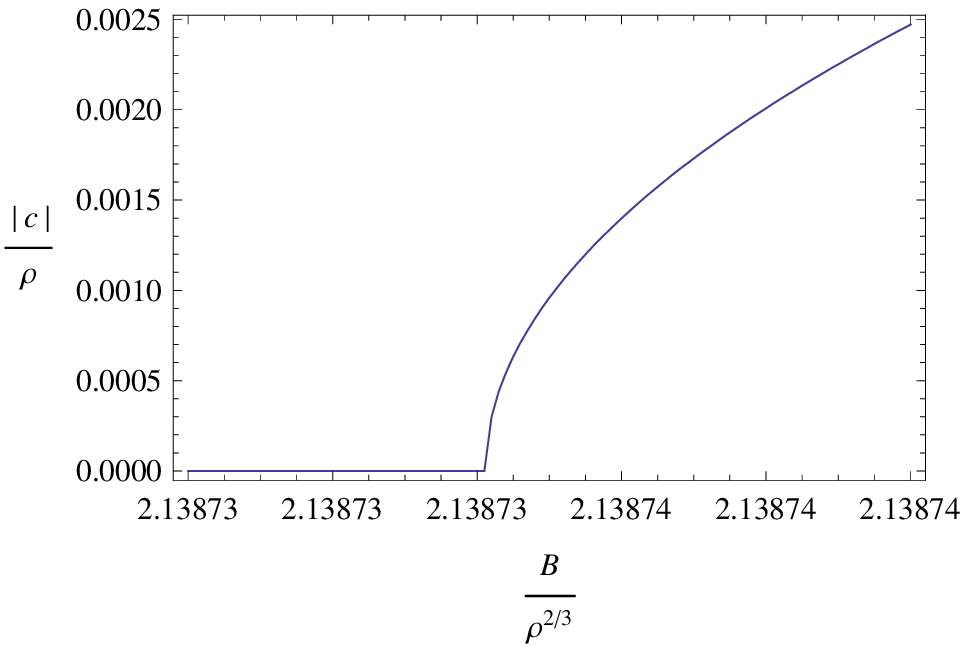} \hspace{.16in}\includegraphics[scale=0.7]{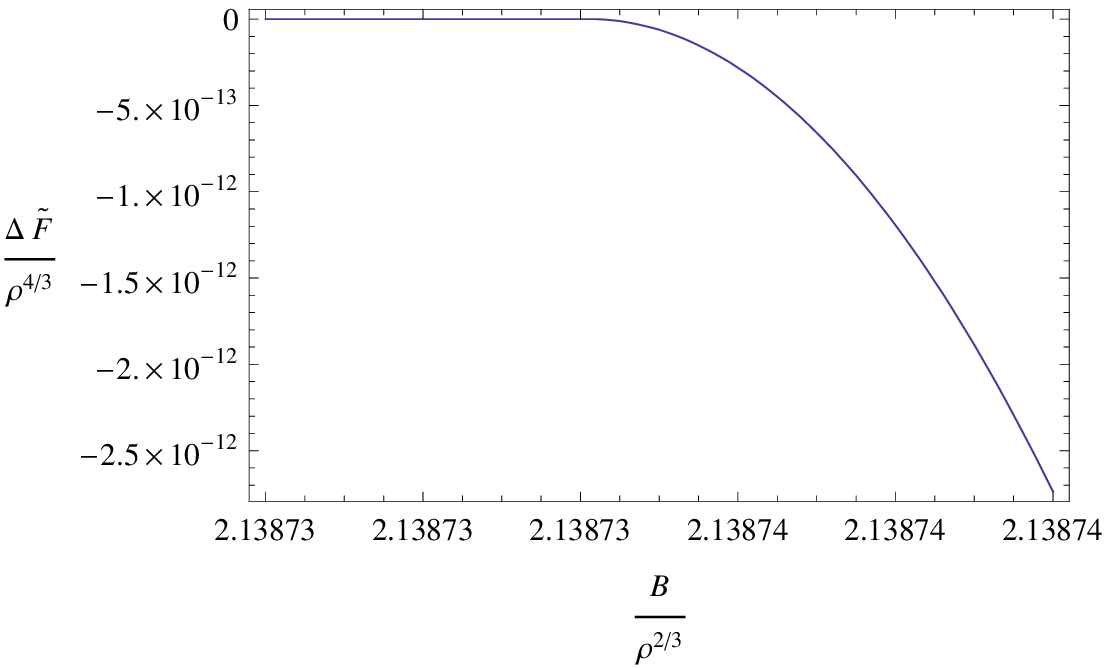}

\caption{ \label{critPlot} On the left, we plot the field theory condensate $c$ as a function of the magnetic field near the critical point.  At small magnetic fields, the condensate vanishes and chiral symmetry is preserved.  Above the transition, a condensate develops and scales as $|c|=1.017(B-B_c)^{1/2}\rho^{2/3}$.  On the right, we plot the difference of free energy density (normalized by 1/$\mathcal{N}$) between the broken and symmetric phases over the same domain.  The broken phase has a lower energy and the difference scales as $\Delta F=-0.07852(B-B_c)^2\mathcal{N}$.}
\end{center}
\end{figure}

The theory therefore has a chiral symmetry breaking transition at $B=B_c$.  Moreover, near the transition, the condensate scales as $c\propto (B-B_c)^{1/2}$ and the difference of free energies as $\Delta F \propto (B-B_c)^2$.  We plot these results in Fig. \ref{critPlot}.  This transition is therefore second order with mean-field exponents.  We have found a concrete realization of a holographic quantum critical point and the primary result of this work.

The mean-field critical exponents indicate that there is a nice Landau-Ginzburg-Wilson effective potential \cite{Wilson:1973jj,SVBSF} in the neighborhood of the critical point.  With the results plotted in Fig. \ref{critPlot}, we can write that effective potential as (in units where $\rho=1$)
\begin{equation}
F_{\rm eff}(c,B)/\mathcal{N}=\alpha_0(B)-\alpha_2(B)(B-B_c)\frac{c^2}{2}+\alpha_4(B) \frac{c^4}{4}+O(c^6),
\end{equation}
where $\alpha_0$ is the free energy of the symmetric embedding and the functions $\alpha_{2,4}$ are numerically determined to be $\alpha_2=0.3038 $ and $\alpha_4=0.2939 $ up to corrections of order $B-B_c$.

We could have also directly measured the effective potential with our brane embeddings.  At fixed magnetic field and density, we can scan the space of all physical values of the mass and condensate by dialing the parameter $\gamma_1$ in the near-horizon series solution Eq. (\ref{horFrob}) for $y$.  Interpreted another way, we can numerically find the expectation value $c$ that corresponds to turning on the mass $m$.  Since the effective potential at nonzero mass is just
\begin{equation}
F_{\text{eff}, m}(c)=F_{\text{eff},m=0}(c)+\mathcal{N}mc,
\end{equation}
the effective potential at zero mass can be found by integrating
\begin{equation}
F_{\text{eff},m=0}'(c)=-\mathcal{N}m.
\end{equation}
This picture only makes sense when the mass (as a function of the condensate) is single-valued.  When the mass is multi-valued, more degrees of freedom are light in the infrared and must therefore be included in the effective potential.

This is exactly what happens when we increase the magnetic field.  For magnetic fields below the critical $B_c$, the mass monotonically decreases with the condensate.  The curve has one zero at zero condensate, corresponding to the symmetric phase.  At the transition, the mass develops a second root at nonzero condensate, but the curve is still single-valued.  This new root is the broken vacuum.  At larger magnetic fields, the curve begins to spiral and at a magnetic field of $B_{\rm spiral}=4.595\rho^{2/3}$ it becomes multi-valued at zero condensate.  The effective potential becomes infinitely curved here.  We conclude that more degrees of freedom are becoming light near the symmetric phase.  On the other hand, the mass curve is single-valued near the second zero and so the effective potential (as a function of only the condensate) seems to be valid near the broken vacuum.

The mass curve continues to spiral as we increase the magnetic field.  In fact, a \emph{second} non-trivial extremum develops at a magnetic field of $B_2=9.664\rho^{2/3}$.  Remarkably, the condensate scales as $c\propto (B-B_2)^{1/2}$ near this new transition.  The new extremum has a lower free energy than the symmetric state with a difference that scales as $\Delta F\propto -(B-B_2)^2$.  Against our intuition, there appears to be a second chiral symmetry breaking transition with mean-field exponents.  We also note that both of these extrema have higher free energy than that of the first non-trivial vacuum.

The simplest way to interpret this result is that a mean-field picture holds near the symmetric extremum (i.e. for small condensates) and the second transition $B=B_2$.  If this is correct, the symmetric extremum must become perturbatively stable (presumably at $B=B_{\rm spiral}$) and then unstable again at $B=B_2$.  This picture would also suggest that the new broken extremum is perturbatively stable and so metastable.  However, we cannot verify any of these claims without computing the fluctuation spectrum around these extrema.

\begin{figure}
\begin{center}

\includegraphics[scale=0.65]{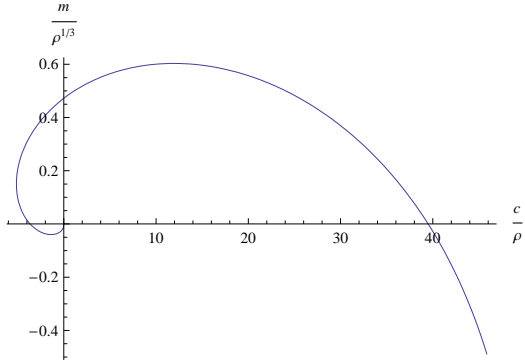}
\caption{\label{spiral} We plot the mass as a function of the condensate at a magnetic field of $B=20\rho^{2/3}$.  The mass has three roots, one at zero condensate and two away from zero.  The three roots correspond to the three extrema of the dual theory at this value of the magnetic field.  The outermost root is the true vacuum of the theory, but we speculate that the other non-trivial root is metastable.
}
\end{center}
\end{figure}

Increasing the magnetic field further, the spiral continues to wind around $(m=0,c=0)$.  We plot the spiral at a magnetic field above the second transition in Fig. \ref{spiral} for a visual aid.  In any event, more and more non-trivial extrema are generated as the spiral winds.  Numerically, there appear to be many chiral symmetry-breaking transitions, each with mean-field exponents.  Finally, if we denote the free energy of the $n^{\rm th}$ non-trivial extremum as $F_n$ (where the actual vacuum is $F_1$) and the free energy of the symmetric extremum as $F_0$, we have
\begin{equation}
F_1 < F_2 < F_3 < \hdots < F_n < F_0,
\end{equation}
at a magnetic field for which there are $n$ non-trivial extrema.  The free energies then form a tower rather than suggesting a landscape.  We speculate that all of the non-trivial extrema are metastable.

\section{The full phase diagram at zero temperature}
\label{fullPhase}
This theory has been studied before at finite density and zero magnetic field \cite{Karch:2007br} as well as at zero density and finite magnetic field \cite{Filev:2007gb}.  We will begin this section by reviewing these works, using their results as a foundation upon which we will construct the phase diagram of this theory at finite density \emph{and} finite magnetic field.  We will find the phase diagram in the grand canonical ensemble, allowing us to find the transition line separating the zero and nonzero density phases.

At zero magnetic field and zero temperature, the theory has two dimensionful parameters - the bare hypermultiplet mass $m$ and the chemical potential $\mu$ - and so the theory can be studied as a function of the dimensionless parameter $m/\mu$.  In \cite{Karch:2007br}, the authors found that there is a second-order transition between a finite density phase and a zero density phase at $m/\mu=1$.  They interpret this result in light of the fact that, at zero density, the lightest charged quasiparticle in the theory has a mass precisely equal to the bare hyper mass $m$.  The field theory interpretation of the transition is simply that a chemical potential greater than that mass is required to condense these charges.

On the gravity side, the finite density embeddings are just as the ones we studied in Sec. \ref{critPoint}: they support nonzero field strength sourced by charge that lies behind the Poinc\'are horizon of AdS.  These embeddings therefore cross that horizon and extend to the bottom of AdS.  Moreover, these horizon-crossing branes only exist for values of chemical potential greater than bare mass, i.e. for $m/\mu \leq 1$.  On the other hand, the zero density embeddings are just those with zero field strength but constant bulk gauge field $A=\mu \,dx^0$.  The equations of motion that govern the brane embedding do not depend on $\mu$ and so these embeddings are the same as at $\mu=0$ \cite{Karch:2002sh}, i.e. constant $y=m$ (with $m$ measured in units of $R$).  The free energy of the zero density embeddings is always zero and the free energy of the finite density embeddings is always negative.  Thus, the finite density phase is always thermodynamically preferred above the zero density phase.  Since that finite density phase does not exist for $m/\mu>1$, there is a transition between the two phases at $m/\mu=1$.  Below the transition, the zero density phase is perturbatively stable (it has the same spectrum as at $\mu=0$ \cite{Kruczenski:2003be}) and so is metastable.  Finally, the mass of the lightest charged quasiparticle in the zero density phase corresponds to the mass of the lightest charged state on the gravity side.  That state is a static fundamental string stretching from the bottom of the brane to the stack of D3 branes at the bottom of AdS.  For a brane with constant $y=m$, that string has a mass $\frac{m}{2\pi\alpha'}$.  In our units, the chemical potential corresponding to this mass is simply $m$, and so a chemical potential larger than this quasiparticle mass indeed condenses charge.

The story is a little less complicated at zero density and finite magnetic field \cite{Filev:2007gb}.  The only dimensionless parameter in the theory is now $B/m^2$ and there is no transition along that line.  The field strength on the brane is a constant $B\, dx^1\wedge dx^2$ and the embedding equations can only be solved numerically.  At all values of $B/m^2$, the D7 brane is repelled from the bottom of AdS, inducing a condensate in the dual field theory.  The theory therefore has a single Goldstone boson (or pseudo-Goldstone boson at finite quark mass) corresponding to the spontaneous breaking of the $U(1)$ chiral symmetry by the condensate.

We now seek to combine these two studies by studying the theory at finite density and finite magnetic field.  We now have three dimensionful parameters - the bare hyper mass $m$, the magnetic field $B$, and the chemical potential $\mu$ - which we combine into the two dimensionless quantities $m/\mu$ and $B/\mu^2$.  We choose to plot the phase diagram as a function of these quantities as shown earlier in Fig. \ref{d3d7phases}.  There are also two order parameters for a transition: the condensate (measured in units of $\mu$) $c/\mu^3$ and the density $d/\mu^3$.  At the least, we should expect a line of finite-density/zero-density transitions connected to the one at zero magnetic field.  There may also be a line of chiral symmetry-breaking/preserving transitions connected to the one at zero mass.

Let us begin mapping out the full phase diagram of this theory in the $\left(\frac{m}{\mu},\frac{B}{\mu^2}\right)$ plane by seeing what happens to the finite/zero density transition away from the limit of zero magnetic field.  As with the transition at $B=0$, the finite density embeddings are the horizon-crossing solutions with nonzero radial electric field and constant magnetic field while the zero density embeddings are the Minkowski embeddings with constant gauge field $A=\mu\, dx^0$ and magnetic field.  These embeddings do not depend on the chemical potential but do depend on the magnetic field.  The mass of the lightest charged quasiparticle is therefore a function of the magnetic field and bare mass through $m_{\rm QP}=m f_{\rm QP}(B/m^2)$.

We then hazard an informed guess about the finite/zero density transition: the chemical potential required to condense charged quasiparticles of some mass should simply be that mass.  Processing this statement, we can define a curve in the $\left(\frac{m}{\mu},\frac{B}{\mu^2}\right)$ plane by
\begin{equation}
\label{zeroDline}
\left(\frac{B}{\mu^2}\right)_{_{\rm crit}} \equiv \left(\frac{m}{\mu}\right)^2 f^{-1}_{\rm QP}\left(\frac{\mu}{m}\right),
\end{equation}
which should give the location of the transition as a function of $m/\mu$.  It is important to remember that, at zero density, the D7 brane is repelled from the bottom of AdS for all values of $B/m^2$.  Thus, the mass of the lightest charged quasiparticle is always greater than the bare mass and so $m/\mu$ is less than or equal to unity along this curve.  Moreover, the lightest charged quasiparticle remains massive as the bare mass goes to zero, so this curve extends all the way to zero mass.  This curve therefore cuts through the $\left(\frac{m}{\mu},\frac{B}{\mu^2}\right)$ plane from one axis to the other.

If the curve Eq. (\ref{zeroDline}) is a transition line of second or higher order (as we might expect it to be, since the zero density phase exists everywhere in the $\left(\frac{m}{\mu},\frac{B}{\mu^2}\right)$ plane but the finite density phase will not), then it is also the spinodal line for the finite density phase\footnote{In this section, we will ignore all of the extra extrema that we found in Sec. \ref{critPoint}.  From our measurements, they have more free energy than the zero density embeddings.}.  That is, we only approach the hypothesized transition line from the finite density side as we take the $d\rightarrow 0$ limit.  This is a tricky numerical limit, since we are taking a parameter to zero but the embedding depends crucially on that small parameter for a large region near the bottom of the brane.   As a result the practical limit is actually that of a large number, namely $B\rightarrow\infty$ (and so $\mu$ and $m\rightarrow\infty$ as well).

\begin{figure}
\begin{center}
\includegraphics[scale=.8]{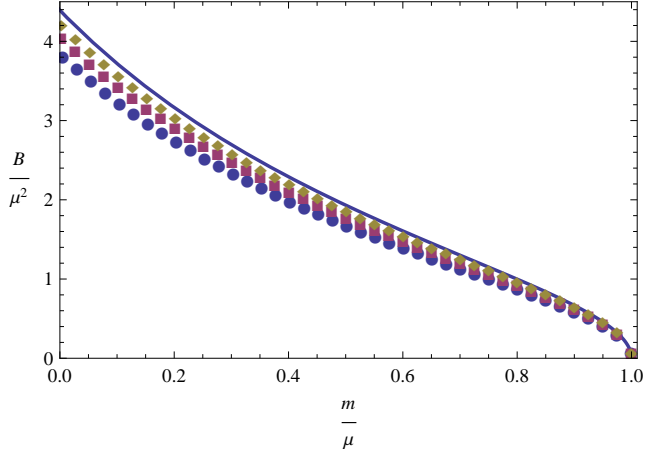}
\caption{ \label{critLimit}
  The transition line between the finite and zero-density phases
  is plotted in blue in the \protect{$\left(\frac{m}{\mu},\frac{B}{\mu^2}\right)$} plane.  The data points indicate finite density extrema at $\rho=1$ and constant magnetic field.  The blue circles are at $B=10$, the purple squares at $B=15$, and the gold diamonds at $B=25$.  As the magnetic field increases, these curves approach the transition line, indicating that the blue curve is indeed the spinodal line of the finite density phase.}
\end{center}
\end{figure}

There is a natural two-step process we use to verify that Eq. (\ref{zeroDline}) is indeed the transition line between the finite and zero density phases of the theory.  First, we verify that this curve is the spinodal line of the finite density phase by studying the $d\rightarrow 0$ limit of the finite density embeddings.  We have plotted the curve Eq. (\ref{zeroDline}) in addition to three lines of data at small density in Fig. \ref{critLimit}.  Each data set is taken at small but fixed density, fixed magnetic field, and varying mass.  As the density decreases (or, the magnetic field increases), each of these curves appears to approach the hypothesized critical line.  We have studied this question further by considering densities as small as $\rho=10^{-6}$ at unit magnetic field and found that the curve Eq. (\ref{zeroDline}) is indeed the spinodal line of the finite density phase within our numerical accuracy.

\begin{figure}

\begin{center}
\includegraphics[scale=.7]{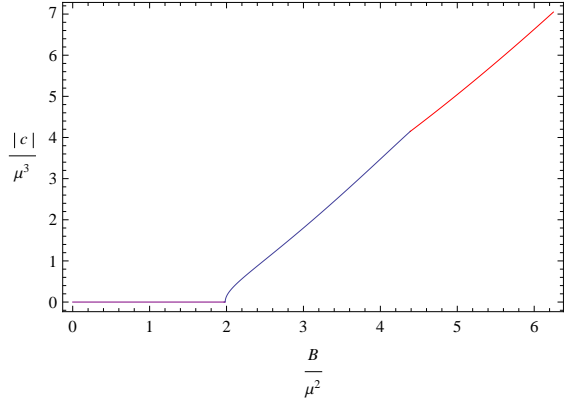}
\caption
  {\label{m=0Condensate} The condensate of the theory at zero mass as a function of the magnetic field.  The purple line indicates the chiral symmetry preserving phase, the blue line the chiral symmetry breaking (but finite density) phase, and the red line the zero density phase.
}
\end{center}
\end{figure}

\begin{figure}

\begin{center}
\includegraphics[scale=.65]{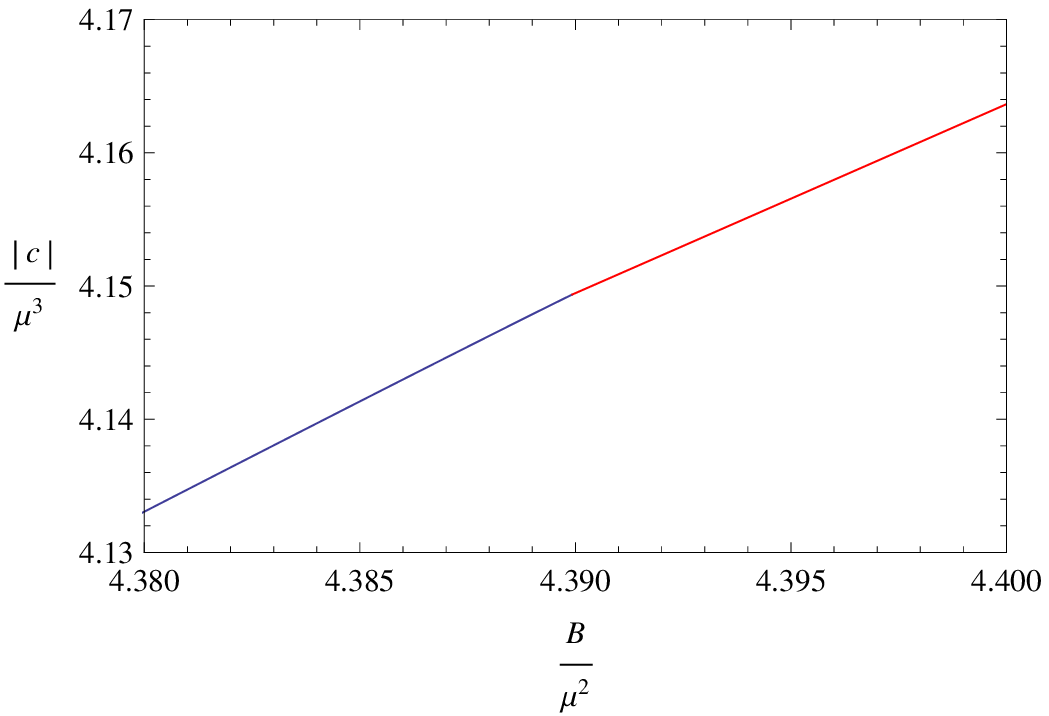}\hspace{.3in}\includegraphics[scale=.65]{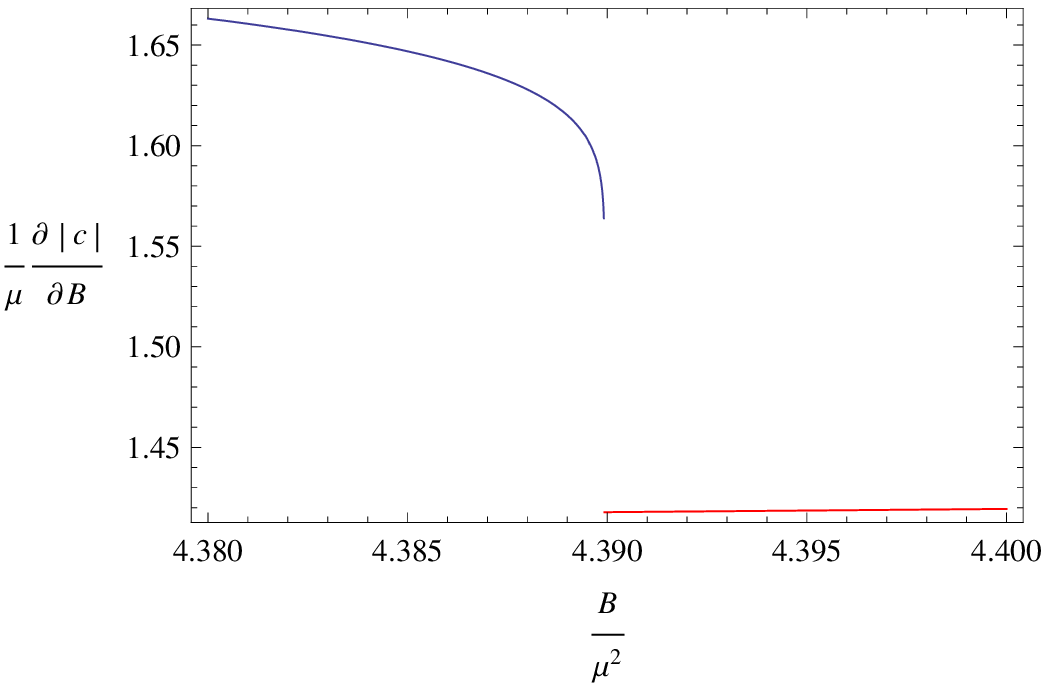}
\caption
  {\label{m=0CondensateZoom} The condensate of the theory (left panel) and derivative of the condensate (right panel) at zero mass as a function of the magnetic field near the finite/zero density transition.  The blue line indicates the chiral symmetry breaking (but finite density) phase and the red line the zero density phase.  Since the condensate appears to be continuous across the transition, the transition must be at least second order.  We lack data for the finite density phase between $B/\mu^2=4.389916$ and the transition at $B/\mu^2=4.389930$, so it is possible that the derivative of the condensate is also continuous at the transition.  If this is so, then the transition is third order.
}
\end{center}
\end{figure}

Next, we should compare the free energy of our finite density embeddings with those of the corresponding zero density embeddings.  Since we also learn about the magnetic properties of the flavor this way, we consider both the free energy and condensate along lines of fixed mass.  For now, we will look at the condensate and save the free energies for our Discussion in Sec. \ref{discuss}.  We begin by plotting the condensate at zero mass in Fig. \ref{m=0Condensate}.  For small magnetic field, the vacuum preserves chiral symmetry and the condensate vanishes; at the critical magnetic field $B_c$, we hit the chiral symmetry breaking transition and the condensate is nonzero.  Next, we zoom in near the spinodal point $B/\mu^2=4.390$ and find that the condensate is continuous between the finite and zero density phases at the spinodal point.  The finite/zero density transition is therefore \emph{at least} second order at zero mass.  We go further and plot the derivative of the condensate in Fig. \ref{m=0CondensateZoom} near the spinodal point.  Unfortunately, we can only approach that spinodal point from the finite density side and so our data only extends to within a small distance of the transition.  Within this domain the derivative of the condensate changes rapidly and may indeed limit to the zero density value at the transition.  The combined results of Fig. \ref{m=0CondensateZoom} therefore indicate that the finite/zero density transition at zero mass is at least second order and may be third order, but no higher.

The last remaining question about the full phase diagram is whether there is a line of transitions connected to the chiral symmetry breaking transition we found earlier at zero mass.  To answer this question, we calculated both the free energy and condensate in the region near the transition in the $\left(\frac{m}{\mu},\frac{B}{\mu^2}\right)$ plane.  We found no evidence of a transition other than at zero mass.  This is consistent with the effective potential picture we developed in Sec. \ref{critPoint}.  Near the critical point, the theory lives at finite density and so we miss no information by switching to the canonical ensemble.  Earlier, we found the effective potential to be
\begin{equation}
F_{\text{eff},m}(c,B)/\mathcal{N}=\alpha_0(B)-\alpha_2(B)(B-B_c)c^2+\alpha_4(B)c^4+mc+O(c^6),
\end{equation}
where we measured $\alpha_{2,4}$ to be constants up to order $B-B_c$ corrections.  For any nonzero mass, this potential exhibits no phase transition near $B=B_c$. There is only an isolated second-order transition in this Landau-Ginzburg-Wilson model corresponding to no mass, just as we see for the D3/D7 system.  For completeness, we should also say that we did not see \emph{any} transitions in our system other than the finite/zero density transition and the chiral symmetry breaking transition at zero mass.

To summarize, the full phase diagram of the theory at zero temperature consists of a single curve Eq. (\ref{zeroDline}) between the finite and zero-density phases.  The transition along this line is at least second and possibly third order along most of the curve.  There is also the second-order transition at zero mass from Sec. \ref{critPoint}.  This transition is isolated.  We therefore arrive at the phase diagram in Fig. \ref{d3d7phases}.  The picture in the canonical ensemble is even clearer.  The zero density phase and the transition to finite density collapses to a line at $d=0$, so that the only interesting feature is the transition at zero mass.

\section{Discussion}
\label{discuss}
\subsection{Quantum Critical Matter}
\label{qcm}
Physical systems with a quantum critical point generally have an anomalous phase of matter.  This phase is found at nonzero temperatures and for values of the control parameters near the critical point.  The anomalous phase is thought to be controlled by the conformal theory at the critical point heated up to a temperature $T$.  In the $(\text{control parameter},T)$ plane, the anomalous phase is roughly wedge-shaped with two borders.  The borders may be continuous transitions or crossovers.  In nature, this ``quantum critical matter'' has been identified by studying both the resistance and the specific heat near the critical point \cite{qCrit}.  For example, the resistance and specific heat at low temperatures in some metals appear to scale with different powers of the temperature inside and outside the wedge.

We would like to identify a quantum critical phase in our theory if it exists.  Following the lead of the condensed matter community, we will search for critical matter by studying the resistance and specific heat near the critical point.  In order to do so, we need to turn on a temperature.  On the gravity side, this amounts to studying probe branes in and supergravity on AdS-Schwarchild$\times S^5$.  Our finite density embeddings still have an electric field supported by charge behind the horizon, so the branes still fall into the black brane of the geometry.  The background metric is now
\begin{equation}
g=\frac{ -f(z) dt^2 + d\vec{x}^2}{z^2} + \frac{dz^2}{f(z) z^2} + d\theta^2 + \cos^2 \theta d\Omega_3^2 + \sin^2 \theta d \phi^2,
\end{equation}
where $f(z) = 1 - \frac{z^4}{z_H^4}$ and $z_H$ is the horizon radius of the geometry, related to the Hawking temperature of the black brane as $z_H = \frac{1}{\pi T}$.  We identify the Hawking temperature with the temperature of the dual theory.  The boundary of the AdS-Schwarzchild space is at $z\rightarrow 0$.  We now wrap probe D7 branes in this geometry as before; the branes wrap a five-cycle inside AdS-Schwarschild as well as the three-sphere, live at constant $\phi$, and are parametrized by the ``slipping mode'' $\theta=\theta(z)$.  In the zero temperature limit, these coordinates are related to our old ones by $y\, z= \sin\theta, r\, z=\cos\theta$.

The DBI action of these probes at fixed density is now given by
\begin{equation}
\hat{S}_{D7}=-\mathcal{N}\int dz \frac{\sqrt{1+z^2f(z)\theta'(z)^2}\sqrt{\rho^2z^6+(1+B^2z^4)\cos^6\theta(r)}}{z^5}.
\end{equation}
We can repeat the analysis of \cite{Karch:2005ms} and holographically renormalize this action.  The counterterms required in this parametrization are a little bit more complicated than those in Sec. \ref{holorg}, but the idea is the same.  The equation of motion for the field $\theta$ has two singular points that concern us: one at the boundary of AdS-Schwarschild and the other at the horizon.  Near the horizon, we simply mandate that $\theta$ is regular.  The series solution for $\theta$ there then has the form
\begin{equation}
\theta(z)=\theta_0+\sum_{i=1}^{\infty}\theta_i\left(\frac{z-z_H}{z_H}\right)^i,
\end{equation}
where the $\theta_i$ are determined by the free parameter $\theta_0$.  Conversely, the near-boundary series solution for $\theta$ is
\begin{equation}
\theta(z)=a_1 z+a_3 z^3+\sum_{i=1}^{\infty}a_{2i+3}z^{2i+3},
\end{equation}
where $a_1$ and $a_3$ are free parameters and the remaining $a_i$ are determined by them.  After performing holographic renormalization, we learn that $a_1$ is the bare hyper mass $m$ and $a_3$ is related to the condensate and mass by $a_3=-\frac{c}{2}+\frac{m^3}{6}$.

The chiral symmetry-preserving embedding is still a solution at nonzero temperature.  In these coordinates, it is simply $\theta=0$.  The broken vacuum, on the other hand, will have $\theta_0\propto (B-B_c(T))^{1/2}$ near the transition.  As before, we elect to shoot in order to find dual extrema at zero bare mass.  We are now set to study the resistance and specific heat.  We first consider the resistance.  The calculation of the resistance of probe brane systems at finite density and magnetic field was first discussed in \cite{O'Bannon:2007in}.  At zero electric field, the DC resistance can be found in terms of one input: the angle at the horizon $\theta_0$.  At first sight, this is promising.  At fixed temperature, $\theta_0$ behaves differently in three distinct regions.  At small magnetic field, chiral symmetry is preserved and $\theta_0=0$; just above the transition, $\theta_0$ scales nicely as $\sqrt{B-B_c(T)}$; finally, at large magnetic field $\theta_0$ simply asymptotes to $\pi/2$.  On these grounds alone, we expect the resistance to behave differently in those three regions.  Indeed, the first two regions are separated by a line of transitions and the last two by a crossover, matching our expectations for a quantum critical phase.  However, it turns out that all of these effects are small.

To see this, consider the resistance at small temperatures (in units where $\rho$ and $\mathcal{N}$ are both unity),
\begin{eqnarray}
\nonumber
\label{resistance}
R_{xx}=R_{yy}&=&\pi^2 T^2+\frac{1}{2}B^2\pi^4\cos^6\theta_0\, T^4+O(T^6), \\
R_{xy}&=& - B + B\pi^6\cos^6\theta_0\, T^6+O(T^8).
\end{eqnarray}
The series expansions Eq. (\ref{resistance}) inform us that the dominant terms in the resistance are uniform near the critical point.  In the diagonal resistance, the first non-trivial difference between the symmetric and broken phases is a term of order $(B-B_c(T))T^4$; in the off-diagonal, the first difference goes like $(B-B_c(T))T^6$.  We therefore conclude that the anomalous resistance of our critical matter is a subleading part of the resistance in both temperature and $B-B_c$.

We move on to consider the specific heat in the neighborhood of the critical point.  First, we compute the specific heat in the symmetric phase.  At zero magnetic field and small temperatures, this scales as a bizarre $T^6$ \cite{Karch:2008fa}.  At nonzero magnetic field, we find that the specific heat scales as $T^2$.  The computation is simple.  Consider the renormalized free energy corresponding to a small temperature $T$ and the symmetric embedding $\theta=0$.  Then we can write the thermodynamic potential of the canonical ensemble as
\begin{equation}
F=-\lim_{\Lambda\rightarrow\infty}\left[\int_{1/\Lambda}^{\infty}dz \hat{\mathcal{L}}[d;0,0]+F_{\rm counter}(\Lambda)\right]+\int_{z_H}^{\infty}\hat{\mathcal{L}}[d;0,0],
\end{equation}
where $F_{\rm counter}(\Lambda)$ is some set of counterterms living on the slice $z=1/\Lambda$.  Only the temperature-dependent part of the potential contributes to the specific heat, so we need only consider the second integral.  For small temperatures, $z_H$ is close to infinity and we can expand the integrand in powers of $z$.  The potential is
\begin{equation}
\frac{F}{\mathcal{N}}=-\pi \rho T-\frac{B^2T^3}{6\rho}+O(T^5),
\end{equation}
and so the entropy density $s=-\left(\frac{\partial F}{\partial T}\right)_{\rho,V}$ is large at zero temperature.  This is a classic feature of probe brane systems and indicates a large ground state degeneracy that is presumably broken away from large $N$ and $\lambda$.  In any event, the specific heat $c_{V}=T\left(\frac{\partial s}{\partial T}\right)_{\rho,V}$ is
\begin{equation}
c_{V,0}=\mathcal{N}\frac{B^2T^2}{\rho}+O(T^4).
\end{equation}

\begin{figure}

\begin{center}
\includegraphics[scale=.6]{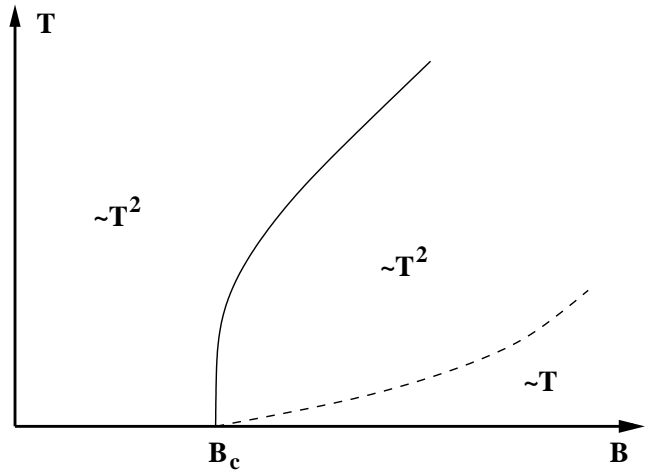}
\caption
  {\label{region} Schematic behavior of the specific heat close to the quantum critical point as a function of temperature. The solid line indicates a line of 2nd order phase transition, the broken line is a crossover as described in the text.
}
\end{center}
\end{figure}

We now endeavor to measure the specific heat in the broken phase.  To do so, we will take advantage of the fact that we have a nice effective potential description of the theory near the critical point,
\begin{equation}
F_{\rm eff}(c,B,T)/\mathcal{N}=\alpha_0(B,T)-\alpha_2(B,T)(B-B_c(T))c^2+\alpha_4(B,T)c^4+O(c^6),
\end{equation}
where everything in sight can be represented as a double series in $T$ and $B-B_c(T)$.  Moreover, we can make the constraint that our theory violates the third law of thermodynamics no worse than in the symmetric phase.  In other words, the zero temperature entropy density is $\pi d$ everywhere.  We have numerically confirmed this result.  The consequence is that neither the coefficients $\alpha_2$, $\alpha_4$, nor the location of the critical magnetic field $B_c$ have a piece linear in the temperature.  At low temperatures, the specific heat is then
\begin{equation}
c_V/\mathcal{N}=-\frac{\alpha_2(B,0)^2B_c''(0)}{2\alpha_4(B,0)}(B-B_c)T+\frac{\alpha_2(B,0)}{2\alpha_4(B,0)}\left(\alpha_2''(B,0)-\frac{\alpha_2(B,0)\alpha_4''(B,0)}{2\alpha_4(B,0)}\right)(B-B_c)^2T+O(T^2),
\end{equation}
where $B_c$ is the location of the critical magnetic field at zero temperature, $B_c(0)$, and $'$ denotes differentiation with respect to temperature.  Before measuring the parameters of the effective potential, we pause to note two things.  First, a wide class of theories with a second order transition and mean-field exponents will have a linear specific heat in the broken phase.  We need not necessarily interpret this result as a signature of fermionic degrees of freedom dominating the broken phase.  Rather, we find a linear specific heat because it is the most generic thing we could find.  Second, in our system chiral symmetry is restored by a temperature, so $B_c(T)>B_c(0)$ and  $B_c''(0)\geq 0$.  However, if $B_c''(0)>0$, the specific heat near the transition at small temperature will be negative.  We therefore predict that $B_c''(0)=0$, a prediction borne out in our numerics.  Indeed, we find that the critical magnetic field behaves as $B_c+a_3 T^3$ at small temperatures.  Moreover, we find that the parameters $\alpha_2''(B)$ and $\alpha_4''(B)$ scale at least as $O(B-B_c)$.  Unfortunately, we have not been able to accurately measure the $O(B-B_c)$ or higher order terms in $\alpha_2''$ or $\alpha_4''$.  It is possible that both of these functions vanish for all $B$, but we have no firm theoretical justification for why they should as we did for the vanishing of $B_c''$ and the linear terms in $T$.  As a result, we expect that there will be a specific heat in the broken phase that scales at least as $(B-B_c)^3T$ at small temperatures.

There is also a large piece of the specific heat that scales as $T^2$.  This is just the specific heat of the symmetric phase.  If the linear piece of the specific heat is $\alpha (B-B_c)^3T$ and the piece from the symmetric phase is $\beta T^2$, then there is a crossover region near the critical point where the specific heat changes its scaling from $T$ to $T^2$.  The line between these two regions is roughly given by $\alpha(B-B_c)^3T_{\rm cross}=\beta T_{\rm cross}^2$, i.e. by $T_{\rm cross}=\frac{\alpha}{\beta}(B-B_c)^3$.  We therefore arrive at a picture for the critical matter in the D3/D7 system, which we present schematically in Fig. \ref{region}. The critical matter is the region in the broken phase where the specific heat scales as $T^2$, rather than as $T$ in the ``normal'' but broken phase below it.

\subsection{Fermions?}
\label{fermions?}
It is tempting to speculate that our quantum phase transition indicates a competition between bosonic and fermionic condensates. While the high density phase is probably dominated by bosons as they do not need
to form a Fermi surface, the chiral symmetry breaking condensate may arise from the interaction of the background magnetic field with the fermion spin.  We might hope that the broken phase is dominated by strongly-coupled fermions.  If so, we would expect to see signs of the de Haas-van Alphen effect.  This effect is a result of the simultaneous presence of Landau levels and a Fermi surface. As the magnetic field is dialed, the spectrum of Landau levels shifts up or down.  For a fixed Fermi surface, there will be particular values of the magnetic field where a Landau level crosses the Fermi surface.  For system of free fermions, these crossings lead to discontinuities and spikes in thermodynamic quantities, including the magnetic susceptibility.  These discontinuities are smoothed out by interactions, but still persist in the form of oscillations or plateaus.

For example, the magnetic susceptibility for a free fermion in a magnetic field will have a series of delta function spikes located at field values
\begin{equation}
B = \frac{ \mu^2 - m^2}{ 2 (l + \frac{1}{2})},
\end{equation}
where $l$ is a positive integer. With interactions, these spikes are smoothed out into a series of oscillations. The ``lowest'' peak will occur for relatively large $B$, and the density of peaks will increase as $B$ is taken to zero.  While the physical interpretation of this effect is most clear at fixed chemical potential, such hills also appear at fixed density.

We have calculated the magnetic susceptibility of our system for magnetic fields above the phase transition and at three values of the mass - the massless case $m = 0$, the relativistic massive case of $m / \mu=1/2$, and the non-relativistic case of $m / \mu=0.999$.  The results are plotted below in Fig. \ref{suscept}.

\begin{figure}

\begin{center}
\includegraphics[scale=0.65]{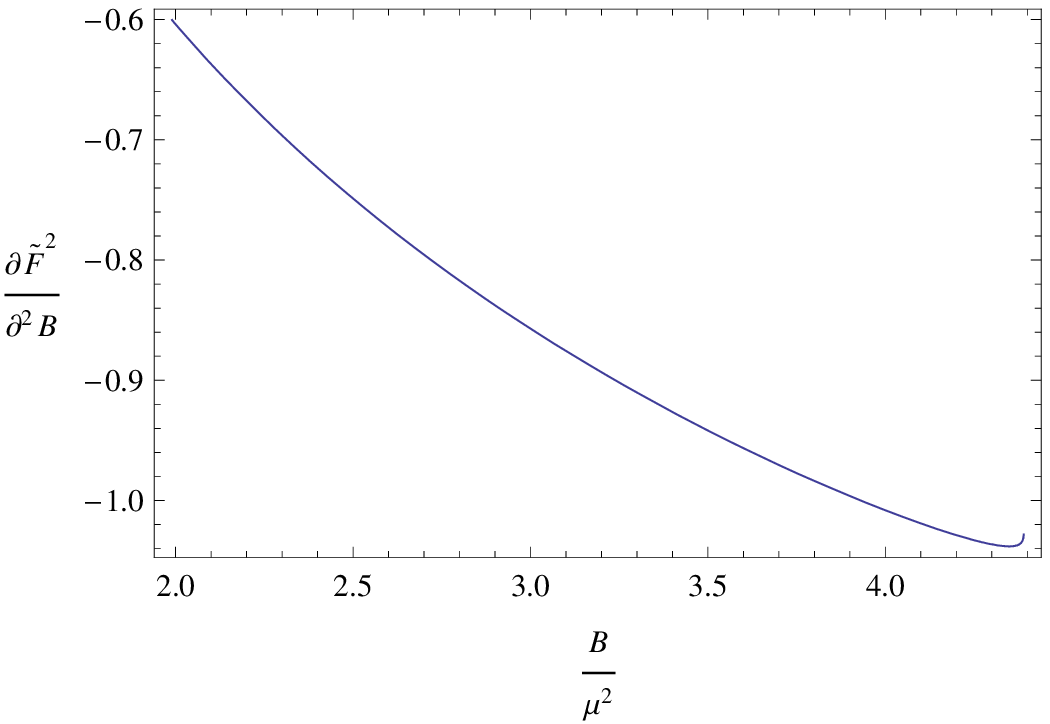}\hspace{.1in}\includegraphics[scale=0.65]{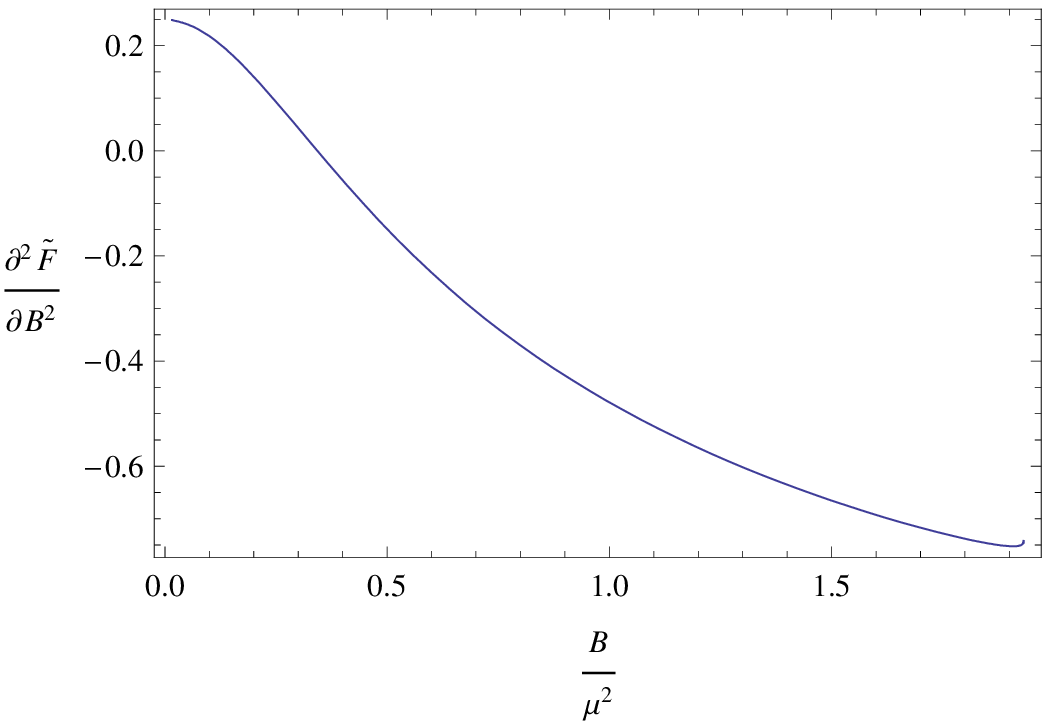}\hspace{.1in}\includegraphics[scale=0.65]{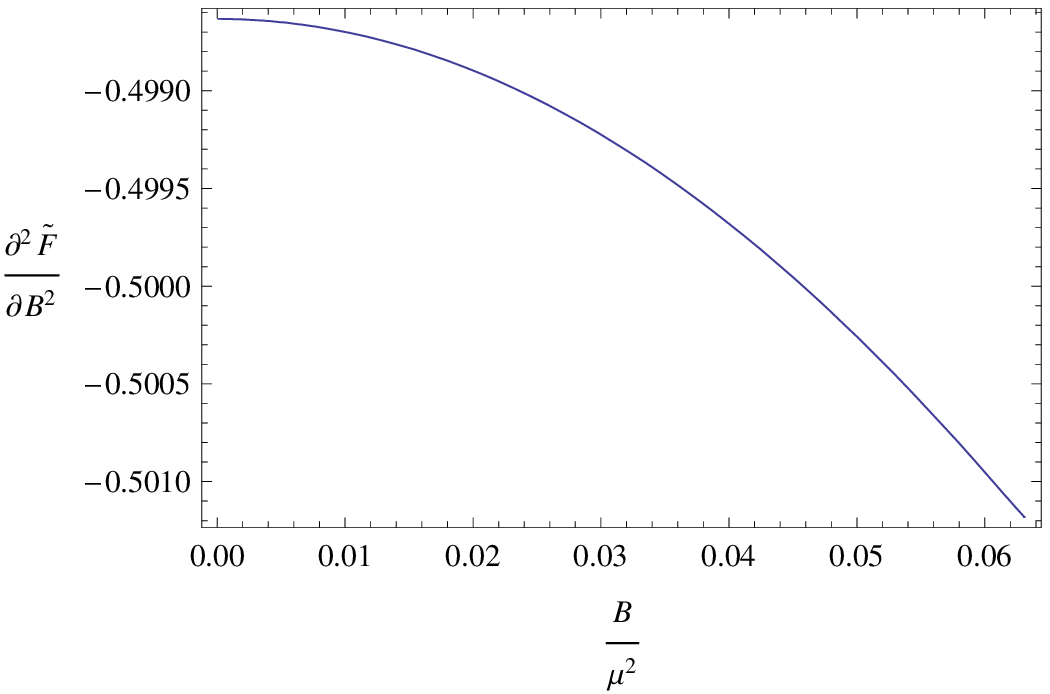}

\caption{\label{suscept} We plot the (normalized) magnetic susceptibility $\mathcal{N}^{-1}\partial^2 F/\partial B^2$ as a function of the magnetic field at $m/\mu=0$ (left), $m/\mu=0.5$ (right), and $m/\mu=0.999$ (bottom).  For the susceptibility at $m=0$, we only plot in the broken phase.  At large magnetic field for both $m=0$ and $m/\mu=0.5$ there is a ``tail'' in the susceptibility.  For $m=0$, the tail is roughly $3\%$ of the total susceptibility and at $m/\mu=0.5$ the tail is smaller, about $1.5\%$.  In each case, we fail to see oscillations in the susceptibility.}
\end{center}
\end{figure}

We do not see a series of oscillations.  The only interesting feature in our susceptibilities is a small upturn at very large magnetic field.  A consistent interpretation of our results at zero mass is that we see the lowest peak of the oscillations in the de Haas-van Alphen effect.  If this is true, then we posit that the higher peaks (occurring at lower magnetic field) are cut out by the critical point.

Within our interpretation, what is happening is that for sufficiently large magnetic fields the system prefers to condense fermions rather than scalars. This is indeed what would happen in our theory at weak coupling and in the non-relativistic regime ($0<\mu -m <<m$). In this weakly-coupled, non-relativistic limit the presence of a magnetic field gives rise to a Zeeman splitting for the spin 1/2 fermions, but not for the spinless scalars. The fermionic state splits into two states, one of which would have a lower energy than the scalar state.  The Zeeman splitting is given by $\Delta E_z = 2\mu_B B$, where $\mu_B$ is the Bohr magneton. For non-relativistic fermions, the splitting is hence of the same size as the spacing between Landau levels, which is also $\frac{ \hbar q B}{mc} =  2\mu_B B$.  By the time the system would be wanting to put fermions in the second Landau level, it has become favorable to once again condense scalars.

It is at least plausible to suspect that the same microscopic physics still dominates our system at strong coupling.
Thus, despite the absence of oscillations in the magnetic susceptibility of our system, we have not been able to rule out the condensation of fermions.

\subsection{A Conjecture}

The quantum phase transition that we have described in this paper for the D3/D7 system arises out of a competition between a finite density and a finite magnetic field.  These competing effects are well understood in the brane picture.  As first established in \cite{Kobayashi:2006sb}, the charged strings required to turn on a nonzero field strength on the branes (and thus a density in the field theory) will pull the probe branes into the horizon, and the only relevant embeddings at finite density are thus the black-hole embeddings.  We see that a finite density attracts the probe branes into the horizon and so tends to restore chiral symmetry.  This result should be equally true in the weak-coupling limit.

On the other hand, a magnetic field on the probes repels them from the horizon \cite{Filev:2007gb}.  For small magnetic fields, this repulsive force is not enough to overcome the attractive force from a finite density, and the resulting embeddings are symmetric.  However, for large magnetic fields the repulsive force is stronger than the attractive, leading to a non-trivial profile for the embedding and thus to the breaking of chiral symmetry.

This scenario is only possible because the dimension of the density operator is always greater than (or equal to, in two spatial dimensions) the dimension of a magnetic field.  The near-horizon dynamics of the D7 branes are then dominated by the finite density.  If, unphysically, the density had a lower dimension than the magnetic field, an infinitesimally small magnetic field would trigger chiral symmetry breaking and there would be no critical point at nonzero magnetic field.

In terms of brane mechanics, our quantum critical point seems as though it could be generic.  Moreover, the fact that we see an analogous transition in the D3/D5 system \cite{D3D5} gives us a little more liberty to speculate.  From our analysis, the necessary conditions for our quantum critical point were (i.) the existence of a global symmetry for which we could turn on a density and magnetic field, (ii.) some chiral symmetry, (iii.) the fact that the magnetic field broke the chiral symmetry at zero density, and (iv.) the conformality of the theory before turning on a density and magnetic field.  We therefore conjecture that all of these conditions are sufficient to guarantee a quantum critical point in the theory at some nonzero magnetic field.  It would be very interesting to investigate the validity of this conjecture in weakly-coupled field theories and for those field theories whose dual is outside the supergravity regime.

\section*{Acknowledgements}
We would like to thank Carlos Hoyos and
Andy O'Bannon for many useful discussions. We would also like to give our special thanks to Piotr Surowka for collaboration during the early stages of this project. This work
was supported in part by the U.S. Department of Energy under Grant Numbers
DE-FG02-96ER40956 and DE-FG02-00ER41132.

\begin{appendix}
\section{Numerical Methods}

This appendix is designed to be a how-to guide for the construction of high-precision numerical brane embeddings.  In it, we will describe the means by which we generate our embeddings as well as how we obtain properties of the dual field theory states.  This will include a discussion of accurate numerical holographic renormalization, as well as the practical implications of the conformal anomaly Eq. (\ref{confAnom}).

We solve for two types of embeddings in this work: (i.) those that fall into the horizon of either AdS or AdS-Schwarschild and (ii.) those that end outside the horizon.  In each case, our solution technique is the same.  The equation of motion for the brane embedding has a regular singular point at the bottom of the brane and at the AdS boundary.  Since we cannot start to numerically integrate the equation of motion at a singular point, we begin by obtaining Frobenius expansions near both the top and the bottom of the brane.  As we mentioned in the text, we elect to shoot from the bottom of the brane.  In practice, this amounts to using the series solution to generate initial conditions a small distance above the bottom.  We then feed those initial conditions into a numerical integrator like Mathematica's NDSolve routine and solve for the embedding up to a close distance below the singular point at the AdS boundary.  We then match the numerically generated solution with the near-boundary series solution to determine the asymptotic data Eq. (\ref{bdyFrob}) of the solution, i.e. the mass and condensate of the dual field theory state.

For completeness, we present the first few terms of the series solutions for the field $y$ in pure AdS and $\theta$ in AdS-Schwarschild at both the top and the bottom of the D7 branes.  First, the near-horizon solution for $y$ in pure AdS is given by
\begin{equation}
y(r)=\gamma_1 r-\frac{\gamma_1^2B^2}{2(1+\gamma_1^2)\rho^2}r^3+\frac{\gamma_1(11+15\gamma_1^2)B^4}{40(1+\gamma_1^2)\rho^4}r^5+O(r^7).
\end{equation}
This series solution obeys the infrared boundary condition $y(0)=0$ and corresponds to finite density embeddings.  The other type of solution is the Minkowski embedding that does not support a density and so ends outside the AdS horizon, i.e. $y(0)\neq 0$.  However, we require the embedding to end smoothly as $\rho\rightarrow 0$ without a conical singularity.  For us this means that the embedding must obey the infrared boundary condition $y'(0)=0$.  The series solution for those embeddings is
\begin{equation}
y(r)=y_0-\frac{B^2}{4y_0(y_0^4+B^2)}r^2+\frac{B^2[(4y_0^4+B^2)^2+4B^2y_0^4]}{64y_0^3(y_0^4+B^2)^3}r^4+O(r^6).
\end{equation}
We implicitly solved for this type of embedding in Sec. \ref{fullPhase}.  In particular, we used these embeddings to obtain the transition line between the finite and zero density phases of the theory.  The parameter $y_0$ is the minimum distance between the brane and the AdS horizon.  It corresponds to the mass of the lightest charged quasiparticle in the field theory, so $y_0=m_{\rm QP}$ for the dual state.  Finally, the solution of $y$ near the AdS boundary takes the form
\begin{equation}
y(r)=m-\frac{c}{2r^2}-\frac{B^2}{4m r^4}+\frac{B^2(2m^3+c)}{8r^6}+O(r^8).
\end{equation}
We found it convenient to compute the near-boundary series solution to very high order, namely $O(r^{-14})$, in order to measure the condensate and free energy to high precision.

The equation of motion and series solutions for $\theta$ in AdS-Schwarschild are much nastier than those for $y$ in pure AdS.  When we measured the chiral symmetry-breaking transition at finite temperature in Sec. \ref{qcm}, we only computed finite density embeddings.  These branes fall into the black hole horizon of AdS-Schwarschild and are regular at the horizon, i.e. they obey the boundary condition $\theta(z_H)=\theta_0$.  The near-horizon solution is then
\begin{equation}
\theta(z)=\theta_0+\frac{3(1+B^2z_H^4)\cos^5\theta_0\sin\,\theta_0}{4(\rho^2z_H^6+(1+B^2z_H^4)\cos^6\theta_0)}\frac{(z-z_H)}{z_H}+O\left(\left(\frac{z-z_H}{z_H}\right)^2\right).
\end{equation}
The near-boundary solution is also more complicated and is given by
\begin{equation}
\theta(z)=mz+\left(-c+\frac{m^3}{3}\right)\frac{z^3}{2}-\frac{m}{40}\left( 10B^2+30mc-3m^4-\frac{5}{z_H^4}  \right)z^5+O(z^7),
\end{equation}
where $m$ and $c$ are the mass and condensate of the dual state.

We used the series solutions above to integrate the equation of motion for $y$ and $\theta$ a small distance away from the bottom of the brane.  For our zero temperature numerics, we computed the near-horizon expansions to $O(r^5)$ and used them to generate initial conditions for $y$ at a position $r=10^{-8}$ for order one values of all other quantities.  At finite temperatures, we usually generated our numerical solutions at a value of $(z_H-z_0)/z_H\sim 10^{-8}$ depending on the temperature (and thus the horizon radius).  Feeding these initial conditions into Mathematica's NDSolve, we would integrate the equation of motion for either $y$ or $\theta$ to values of either $r_{\rm max}\sim 10^9$ or $z_{\rm min}\sim 10^{-9}$.  We also drove up the precision and accuracy goals inside of NDSolve depending on the context.  For the numerics near the transition, we usually turned the working precision of all our numerics up to $17-18$.  Away from the transition, we usually found that we needed higher working precisions in order to ensure stable numerical solutions and accurate measurements of field theory quantities.

Once we generated an embedding, we characterized it with its asymptotic data.  We have found that an accurate way to measure that data is to perform a two-step process.  First, we construct a table of the form $\{r,y(r)\}$ or $\{z,\theta'(z)\}$ very close to the boundary.  We then fit this table to measure the mass of the dual state $m=y(0)$ or $m=\theta'(0)$.  Having fit the mass, we obtain the condensate by matching the embedding with the near-boundary series solution much further away from the boundary.  At zero temperature, for example, we matched at a value of $r=20$ for embeddings ``close'' to the critical point.  If we match much closer to the boundary, the subleading term that contains information about the condensate dies too quickly and the numerical matching is unstable.

The next step in our analysis depended on our objective.  When we found the critical point in Sec. \ref{critPoint}, we varied the initial parameter $\gamma_0$ in the near-horizon solution in order to locate the embeddings with asymptotics $m=0$.  However, when we measured the magnetic susceptibility at fixed mass in Sec. \ref{fermions?} we next obtained the chemical potential $\mu$ by integrating
\begin{equation}
\mu=\int_0^\infty dr A'_0(d;y(r),y'(r))
\end{equation}
along the embedding.  We then varied $\gamma_0$ until we found the embeddings with $m/\mu=$constant.

The last quantity we measured for an embedding is also the most difficult to measure.  The renormalized free energy Eq. (\ref{renS}) corresponding to a brane solution is obtained by some numerical trickery that we will now discuss.  Recall that we obtained the renormalized free energy by beginning with the (divergent) bulk action, cutting off the integration a short distance away from the boundary, adding counterterms on that radial slice, and then removing the cutoff.  This is a numerically intractable procedure: it is the numerical equivalent of subtracting $\infty-\infty$.  Instead, we modify the bulk action by subtracting the appropriate terms.  The radial integral of these terms up to the cutoff simply produces the slice counterterms.  The modified Lagrangian is convergent near the boundary and so suitable for numerical integration.  At zero temperature, for example, our modified bulk action is
\begin{equation}
\label{newS}
\hat{S}_{D7,new}(\Lambda)=-\mathcal{N}\int^\Lambda_1 dr\left[\hat{\mathcal{L}}_{D7}[d;y(r),y'(r)]-r^3-\frac{B^2}{2r}\right]-\mathcal{N}\int^1_0dr\left[\hat{\mathcal{L}}_{D7}[d;y(r),y'(r)]-r^3\right],
\end{equation}
where we broke the integral up into two parts so that the integral of the $1/r$ term gives the logarithmic counterterm.  As we take the $\Lambda\rightarrow\infty$ limit, this modified action converges to the renormalized one.  However, our story does not end here.  The modified Lagrangian $\hat{\mathcal{L}}-r^3-B^2/2r$ tends to behave poorly for numerical solutions at large $r$.  We solve this problem by breaking up the first integral in Eq. (\ref{newS}) into a numeric part and an analytic part.  The numeric part is simply the first integral up to a cutoff where we matched the condensate to the embedding.  We find the analytic part by series expanding the integrand of the first integral in powers of $1/r$ to high order in terms of the asymptotic data $(m,c)$ of an arbitrary embedding.  Our numerical holographically renormalized action then takes the form
\begin{equation}
\hat{S}_{D7,ren}=-\mathcal{N}\int_\Lambda^{\infty} \mathcal{L}_{\rm analytic}(d,m,c)+\hat{S}_{D7,new}(\Lambda),
\end{equation}
where $m$ and $c$ take on their measured values.

We find that a good test of all our numerics is to compute the difference of free energies between the trivial embedding $y=0$ and one arbitrarily close to it.  If we do our job right, this difference vanishes as we let the numerical embedding become arbitrarily close to the trivial one.  Moreover, each piece of the numerics must function correctly to pass this test.  By adjusting the parameters we mentioned above and not working too hard, we were able to resolve differences of free energy at zero temperature down to $10^{-20}$, much smaller than we needed to verify any of the results in this work.  The finite temperature numerics are more temperamental: near the critical point, we resolved differences of order $10^{-14}$.

Finally, we want to mention the practical consequences of the conformal anomaly Eq. (\ref{confAnom}).  As we discussed in the text, the simplest way to find a brane embedding is to solve an equation of motion at fixed density.  We are then naturally in the canonical ensemble.  The chemical potential is determined by the density and the embedding.  In order to transform our results into statements in the grand canonical ensemble, we must find a class of embeddings at fixed chemical potential.  One way to do this is to generate a host of embeddings at varying density and shoot for the desired chemical potential.  This is silly; instead, it is easier to obtain results at fixed density, say $\rho=1$, and then use dimensional analysis to find the corresponding result at fixed $\mu$.  Since dimensional analysis is actually a dilatation, the free energy picks up a logarithmic term from the conformal anomaly.  For example, say that we have an embedding with some chemical potential $\mu$.  We want to find the free energy of the related embedding at $\mu=\mu_0$; that embedding is related to the original one by a scale factor $\lambda=\mu_0/\mu$. The free energy density of the new embedding is
\begin{equation}
F'=\frac{\mu_0^4\left(F+\frac{B^2}{2}\log\left( \frac{\mu}{\mu_0}\right)\right)}{\mu^4},
\end{equation}
where $F$ is the free energy density of the original embedding.  The first term is the one you get from naive dimensional analysis and the second (proportional to $\hbar$) is the anomalous contribution.  We employed this procedure throughout Sec. \ref{fullPhase} and \ref{discuss}, collecting data at $\rho=1$ and then translating it into data at $\mu=1$.
\end{appendix}

\bibliography{bd}
\bibliographystyle{JHEP}

\end{document}